\documentclass{aastex61}
\usepackage[T1]{fontenc}
\usepackage{graphicx,aas_macros}
\usepackage{setspace}

\def\kms{\rm km/s}

\def\kms{{\rm km\:s^{-1}}}

\begin{document}

\title{On the connection between turbulent motions and particle acceleration in galaxy clusters}
\shorttitle{Turbulence to radio power connection in galaxy clusters} \shortauthors{D.~Eckert et al.}

\author{D.~Eckert} 
\altaffiliation{E-mail: Dominique.Eckert@unige.ch}
\affiliation{Department of Astronomy, University of Geneva, 16, ch. d'Ecogia, CH-1290 Versoix, Switzerland}
\author{M.~Gaspari}
\altaffiliation{{\it Einstein} and {\it Spitzer} fellow}
\affiliation{Department of Astrophysical Sciences, Princeton University, 4 Ivy Lane, Princeton, NJ 08544-1001 USA}
\author{F.~Vazza}
\affiliation{INAF - Istituto di Radioastronomia, via Gobetti 101, 40129, Bologna, Italy}
\author{F.~Gastaldello}
\affiliation{INAF - IASF-Milano, Via E. Bassini 15, 20133 Milano, Italy}
\author{A. Tramacere}
\affiliation{Department of Astronomy, University of Geneva, 16, ch. d'Ecogia, CH-1290 Versoix, Switzerland}
\author{S. Zimmer}
\affiliation{DPNC, University of Geneva, 24, Quai Ernest-Ansermet, CH-1211 Geneva, Switzerland}
\author{S.Ettori}
\affiliation{INAF - Osservatorio Astronomico di Bologna, Via Pietro Gobetti 93/3, 40129 Bologna, Italy}
\affiliation{INFN, Sezione di Bologna, viale Berti Pichat 6/2, 40127 Bologna, Italy}
\author{S. Paltani}
\affiliation{Department of Astronomy, University of Geneva, 16, ch. d'Ecogia, CH-1290 Versoix, Switzerland}



\begin{abstract}
\noindent
Giant radio halos are Mpc-scale diffuse radio sources associated with the central regions of galaxy clusters. The most promising scenario to explain the origin of these sources is that of turbulent re-acceleration, in which MeV electrons injected throughout the formation history of galaxy clusters are accelerated to higher energies by turbulent motions mostly induced by cluster mergers. In this Letter, we use the amplitude of density fluctuations in the intracluster medium as a proxy for the turbulent velocity and apply this technique to a sample of 51 clusters with available radio data. Our results indicate a segregation in the turbulent velocity of radio halo and radio quiet clusters, with the turbulent velocity of the former being on average higher by about a factor of two. The velocity dispersion recovered with this technique correlates with the measured radio power through the relation $P_{\rm radio}\propto\sigma_v^{3.3\pm0.7}$, which implies that the radio power is nearly proportional to the turbulent energy rate. Our results provide an observational confirmation of a key prediction of the turbulent re-acceleration model and possibly shed light on the origin of radio halos.\\
\end{abstract}

\keywords{acceleration of particles, radiation mechanisms: nonthermal, radio continuum: general, turbulence, galaxies: clusters: general}

\section{Introduction} \label{s:intro}
\noindent
Giant radio halos are Mpc-scale diffuse radio sources associated with the central regions of galaxy clusters \citep[see][for a review]{feretti12}. Radio halos are a transient phenomenon observed in only a fraction of galaxy clusters \citep[e.g.][]{venturi07,venturi08,basu12} and recent studies have shown that they occur only in dynamically disturbed systems \citep{buote01,cassano10,rossetti11}, suggesting a connection between particle acceleration and major cluster mergers. 

The kinetic energy injected during major mergers first generates a turbulent cascade down to small spatial scales \citep{dolag05,vazza09b,vazza17,miniati15} which may ultimately be dissipated into gas heating, magnetic field amplification and cosmic-ray acceleration. Stochastic acceleration generated by turbulent motions in the intracluster medium (ICM) is now thought to be the most plausible particle acceleration mechanism to re-energize a pre-existing population of $\sim$MeV electrons and explain the origin of radio halos \citep[e.g.,][]{brunetti01,brunetti11,brunetti16,petrosian08}. The turbulent re-acceleration model reproduces a number of observed features, such as the existence of steep-spectrum radio halos \citep{brunetti08,macario10}, the curvature of the spectra at high frequency \citep{thierbach03} and the link between radio halos and cluster mergers \citep{cassano10}. However, the connection between radio halos and ICM turbulence remains untested.

In this Letter, we use the power spectrum of X-ray surface-brightness fluctuations in the ICM to search for a connection between turbulent motions and radio properties in galaxy clusters. Recent theoretical progress \citep{schuecker04,churazov12,gaspari13,gaspari14,zhu14} has shown that gas density fluctuations act as a passive tracer of velocity fluctuations in the ICM and that the maximal amplitude of density fluctuations is linearly related to the turbulent Mach number. This method has been successfully applied to several clusters thus far \citep{zhu15,hofmann16,arevalo16,khatri16} and it was found to reproduce the level of turbulence directly measured in the Perseus cluster by \emph{Hitomi} \citep{hitomiturb}. We retrieve the amplitude of density fluctuations at a fixed spatial scale and use this quantity as a proxy for the expected level of turbulence. We then search for a connection between turbulence and radio emission. The paper is organized as follows. In \S\ref{s:analysis}, we present the sample and describe the methodology adopted to analyze the X-ray data. In \S\ref{s:results}, we present our results and discuss them in the framework of the turbulent re-acceleration model in \S\ref{s:interp}. In \S\ref{s:conc}, we summarize our main conclusions.

\section{Analysis} \label{s:analysis}
\subsection{The sample} \label{s:sample}
\noindent

We base our analysis on the sample of 55 clusters with available radio information at nearly uniform depth from \citet{cassano13}. The sample is based on the GMRT radio halo survey \citep{venturi07,venturi08} with the addition of clusters with known radio emission from the literature. We searched the \emph{XMM-Newton}, \emph{Chandra} and \emph{ROSAT}/PSPC archives for available X-ray data and selected a subsample of 51 clusters for which the quality of the X-ray data is sufficient to retrieve the level of surface brightness fluctuation over large spatial scales. A radio halo was detected for 25 clusters in our sample, whereas for the remaining 26 systems upper limits to the radio flux at 1.4 GHz are available. For nearby clusters ($z<0.1$), we use \emph{ROSAT}/PSPC, as the wide field of view (FOV) of this instrument allows us to cover uniformly a circular region of 1 Mpc radius, i.e., comparable to the typical sizes of radio halos. In the redshift range $0.1<z<0.3$ we use \emph{XMM-Newton} as our instrument of choice given its large collecting area and FOV, with the exception of a few cases (e.g., Bullet, A520) for which deep \emph{Chandra} data are available. Beyond $z=0.3$ only the angular resolution of \emph{Chandra} is sufficient to resolve scales less than $\sim50$ kpc, thus we restrict ourselves to \emph{Chandra} data. The final sample is presented in Table \ref{tab:prop} together with its relevant properties and the adopted instrument. In all cases, we restrict ourselves to the [0.5-2] keV band to be sensitive only to density fluctuations. Given that our sample comprises only hot clusters this choice of energy band is appropriate.

\subsection{Data reduction}
\noindent

\subsubsection{XMM-Newton}
\noindent
We reduced the \emph{XMM-Newton} data using XMMSAS v15.0 and the ESAS software package \citep{snowden08}. Time periods including flaring soft proton flux are filtered out to extract clean event files. We use the unexposed corners of the CCD chips to measure the quiescent background level in each observation. We then renormalize the filter-wheel-closed datasets to match the count rates measured in the CCD corners, which allows us to create model particle background images for each observation. We extract photon images and exposure maps in the [0.5-2] keV band from the cleaned event files. Finally, to avoid contamination from point sources we run \texttt{ewavelet} in each observation and mask a circular region of 30 arcsec radius around each point source. For more details we refer the reader to \citet{eckert14}.

\subsubsection{Chandra}
\noindent
We analyzed the data using the CIAO v4.9 software package and CALDB v4.7.3. For each observation, the raw data are reprocessed with the latest calibration files by running the \texttt{chandra\_repro} pipeline. Periods of flaring background are removed by using the \texttt{deflare} tool. We then extract photon images and exposure maps in the [0.5-2] keV band. To estimate the local background, we use the \texttt{blanksky} and \texttt{blanksky\_image} tools \citep{HM06} to estimate the count rate in the [9.5-12] keV band and renormalize blank-sky datasets to the appropriate level for each observation. Point sources are detected using \texttt{wavdetect} and masked during the analysis.

\subsubsection{ROSAT/PSPC}
\noindent
We reduced the \emph{ROSAT} data using the Extended Source Analysis Software package \citep{snowden94}. The analysis pipeline follows exactly the method presented in \citet{e12}. We create photon images, exposure maps and background in the \emph{ROSAT} R37 band (corresponding to [0.42-2.01] keV) and mask the detected point sources according to the size of the PSF at each radius. For more details on the analysis procedure we refer to \citet{e12}. 

\subsection{Temperature measurements}

We extracted mean spectroscopic temperatures within the same region of 1 Mpc radius as for the extraction of the amplitude of surface-brightness fluctuations. For \emph{XMM-Newton}, we follow the method outlined in \citet{eckert14}. Briefly, we use a phenomenological model to describe the spectral shape of the non X-ray background, which we fit together with the source. The sky background is modeled as the sum of two APEC models \citep{smith01xray} for the Galactic foregrounds and an absorbed power law for the cosmic X-ray background. In the case of \emph{Chandra}, we use an offset region to describe the local background. In both cases, the source is described as a single-temperature APEC model absorbed by the Galactic $N_{H}$, which we fix to the 21cm value \citep{kalberla05}. The spectra are then fit in {\sc Xspec} using C statistic \citep{cash79}. The best-fit spectroscopic temperatures are reported in Table \ref{tab:prop}.

\subsection{Surface brightness fluctuations}
\noindent
To compute the amplitude of density fluctuations, we create a two-dimensional model for the large-scale gas distribution and extract a residual map by dividing the observed emission by the model brightness. The amplitude of surface-brightness fluctuations is then recovered from the Fourier power spectrum of the fluctuations in the residual image at the desired scale. 

To model the cluster gas distribution, we assume that the morphology of the cluster can be described by an elliptical beta model \citep{cavaliere77}. We use a weighted principal component analysis method to determine the centroid of emission and the ellipticity parameters (major and minor axes, rotation angle). We mask obvious substructures associated with individual sub-clumps (e.g. the Bullet in 1E 0657-56) to avoid introducing additional power unrelated to turbulent motions. We then extract a surface brightness profile using {\sc Proffit} \citep{eckert11}, and we fit the beta model to the data using C statistic. In the cases where a double beta model is statistically favored we adopt the double beta model solution as our model of choice. We then create a model image for the two-dimensional gas distribution by folding the best-fit model with the exposure map and adding the background map. 

To compute the Fourier power spectrum at the desired scale, we use the modified $\Delta$-variance method introduced by \citet{arevalo12}, which allows to take the presence of holes and non-periodic boxes into account. In this method, the image and the mask are convolved with a mexican hat filter and the filtered image is corrected for the spurious features introduced by the presence of holes. The variance of the filtered image is proportional to the power at the chosen scale. To estimate the noise level, we simulate Poisson noise on top of the model image and apply the same procedure. The variance of the noise image is then subtracted from that of the true image. The uncertainty in the measurement of the power spectrum is estimated by splitting the filtered image into 20 subregions, computing the power in each region separately, and performing $10^4$ bootstrap resampling of the measured values. We then adopt the standard deviation of the bootstrap distribution as our $1\sigma$ error. 

\subsection{Turbulent velocity}
\label{sec:turb}

As shown by \citet{arevalo12}, filtering the image at a given scale $\ell$ allows to select fluctuations at a wave number $k=\frac{1}{\sqrt{2\pi^2}\ell}$. For the present work, we choose a smoothing scale $\ell=150$ kpc, corresponding to $k^{-1}=660$ kpc. Such a scale roughly corresponds to the typical size of groups accreting onto clusters, i.e. it should be close to the injection scale of turbulent motions, at which the amplitude of fluctuations is expected to peak. Finally, the projected two-dimensional amplitude is converted into three-dimensional fluctuations $\delta\rho/\rho$ by using the best-fit beta model parameters and computing numerically the power induced by the projection of the emissivity distribution along the line of sight \citep[see Eq. 11 of][]{churazov12}. For consistency, we checked that the amplitudes recovered from {\it Chandra} and \emph{XMM-Newton} data are in agreement when data from both telescopes are available. We found an excellent agreement between the two X-ray telescopes, giving us confidence that our method is robust.

As shown in \citet{gaspari13}, in the subsonic regime the maximum amplitude of density fluctuations is linearly related to the turbulent Mach number $M=\frac{\sigma_v}{c_s}$, with $\sigma_v$ the turbulent velocity dispersion and $c_s$ the sound speed in the medium. The relation reads $M_{\rm 1D}\approx2.3\,\frac{\delta\rho}{\rho}$ or $M_{\rm 3D}\approx4\frac{\delta\rho}{\rho}(L/{\rm 500 kpc})^{-0.25}\approx3.7\frac{\delta\rho}{\rho}$ for our choice of scale. The intrinsic scatter of the relation is expected to be $\sim30\%$ \citep{zhu14}. For each cluster, we compute the average spectroscopic temperature within the same circular region of 1 Mpc radius to estimate the average sound speed $c_s=(\gamma kT/\mu m_p)^{1/2}$ with $\gamma=5/3$ and convert the Mach number into turbulent velocity.

\section{Results} \label{s:results}

\subsection{Amplitude of surface-brightness fluctuations} \label{sec:2d}
\begin{figure*}
\resizebox{\hsize}{!}{\hbox{\includegraphics{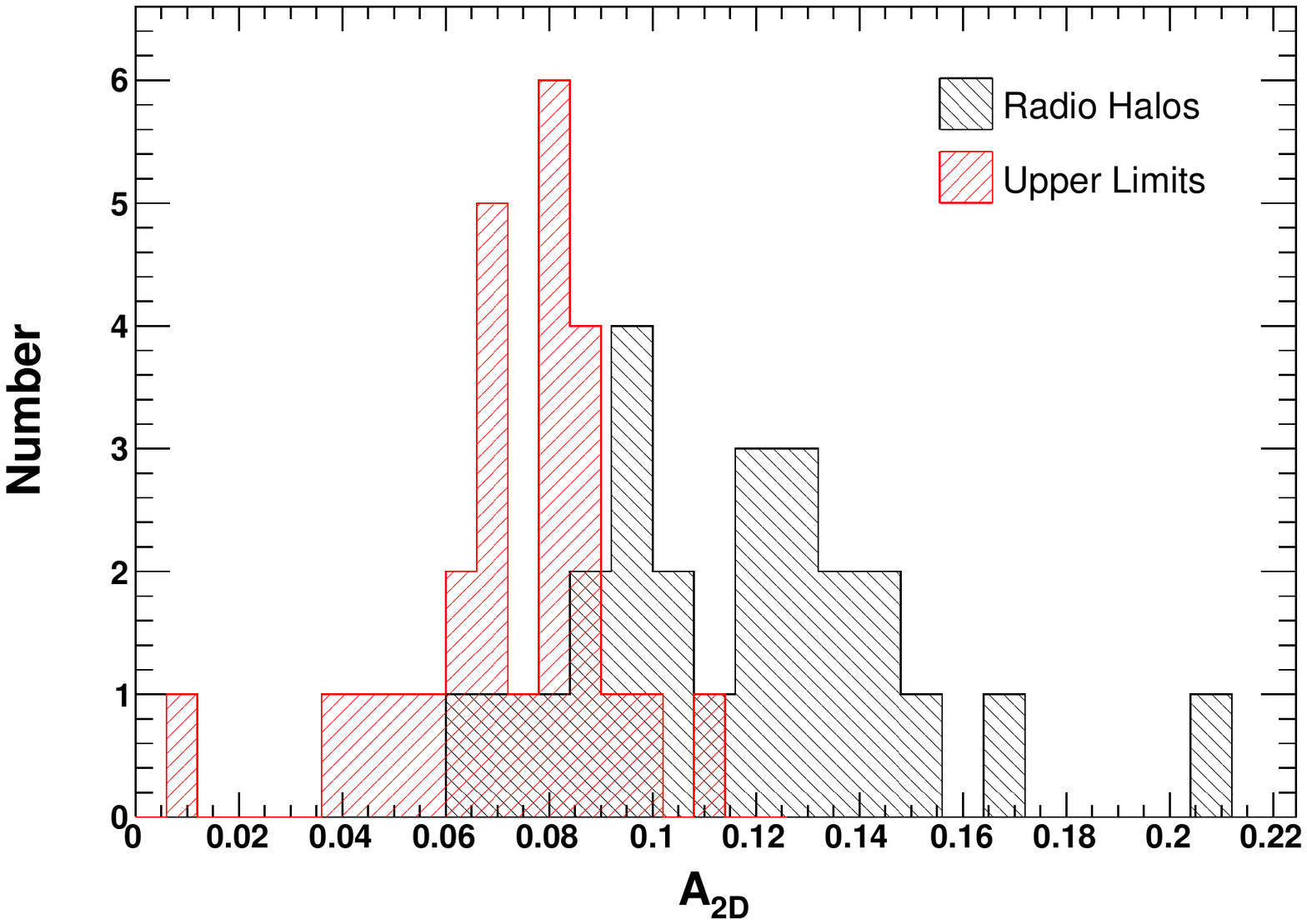}\includegraphics{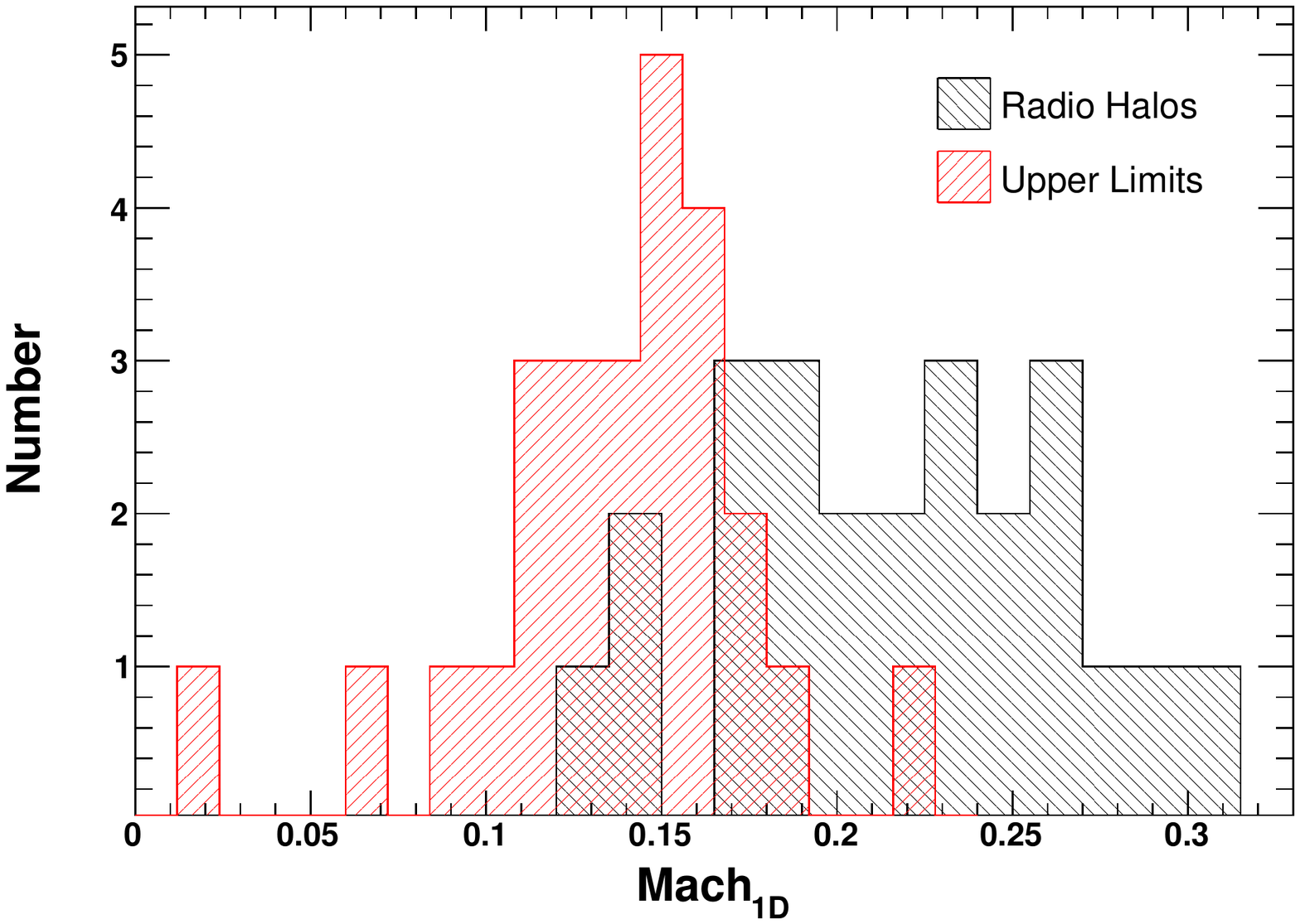}}}
\caption{\emph{Left:} Distribution of two-dimensional fractional amplitude of surface-brightness fluctuations for radio halo clusters (black) and systems without a radio halo (red). \emph{Right:} Same as in the left panel for the 1D turbulent Mach number estimated from the deprojected density fluctuations.}
\label{fig:distribution}
\end{figure*}

We estimated the two-dimensional amplitude of surface-brightness fluctuations $A_{2D}$ at a fixed scale $k^{-1}=660$ kpc and the corresponding Mach number and turbulent velocity dispersion for all clusters in our sample. The resulting values are provided in Table \ref{tab:prop}. In the left-hand panel of Fig. \ref{fig:distribution} we show the distribution of the values of $A_{2D}(k=1/660\mbox{ kpc})$ split between the radio halo and non radio halo populations. We see that the radio halo clusters exhibit on average a higher value of $A_{2D}$ than the upper limit population, with mean values $A_{2D,RH}=0.117$ and $A_{2D,UL}=0.072$ for the radio halo and upper limit populations, respectively. We performed a Kolmogorov-Smirnov (K-S) test to determine the probability that the two sets of values are drawn from a common parent population. The K-S test returns a probability of $4\times10^{-6}$, meaning that the two distributions are different at $4.6\sigma$ confidence. 

Using the deprojection factors estimated in \S\ref{sec:turb}, we then computed the amplitudes of 3D fluctuations of $\delta\rho/\rho$, which we converted into 1D Mach numbers using the relation $M_{\rm 1D}\approx2.3\,\delta\rho/\rho$ \citep{gaspari13}. The distribution of the values of $M_{\rm 1D}$ is shown in the right-hand panel of Fig. \ref{fig:distribution}, again splitting the sample into the radio halo and radio quiet populations. A clear segregation is observed between the two populations, most radio halo clusters exhibiting a value of $M_{\rm 1D}$ in the range  0.15-0.3 (which is consistent with the sample of \citealt{hofmann16}), whereas for the clusters where no radio emission was detected, the typical value of $M_{\rm 1D}$ is $\sim0.12$ and always less than $\sim0.2$. In this case, the K-S test returns a probability $p=2\times10^{-7}$ that the two datasets are drawn from the same parent distribution, i.e. the result is significant at $5.2\sigma$.

\subsection{Correlation with radio power}

\begin{figure*}
\centering
\resizebox{0.8\hsize}{!}{\includegraphics{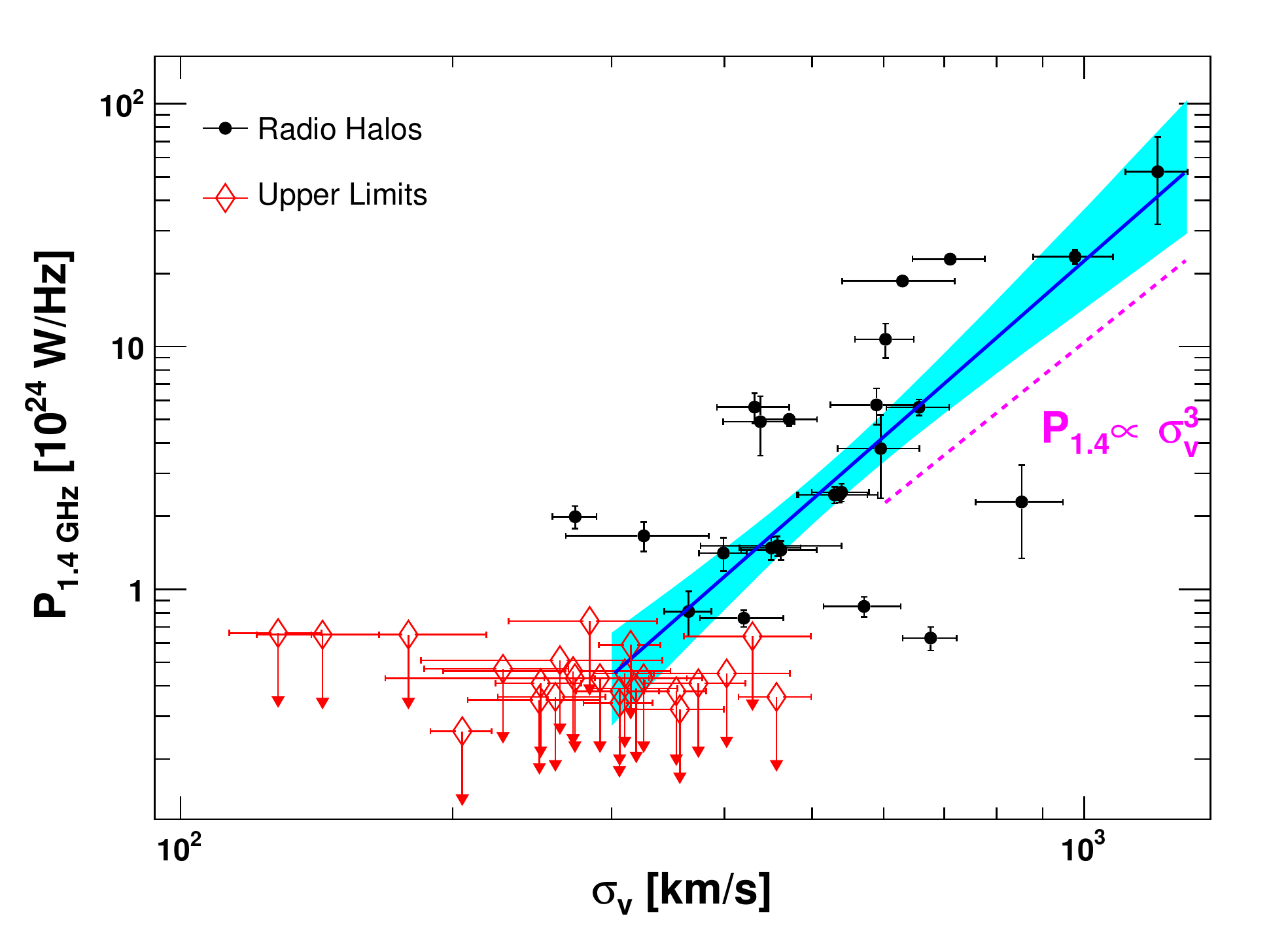}}
\caption{Radio power at 1.4 GHz as a function of velocity dispersion $\sigma_v=M_{\rm 3D}\,c_{\rm s}$ for radio halo clusters (black) and radio upper limits (red). The blue curve and shaded area show the best fit to the data with a power law and its $1\sigma$ error envelope. }
\label{f:a2d_data}
\end{figure*}

For each cluster, we used the average temperatures reported in Table \ref{tab:prop} to determine the sound speed in the medium. We then estimated the average three-dimensional turbulent velocity dispersion through the relation $\sigma_{v}=M_{\rm 3D}\,c_{\rm s}\approx3.7\,c_{\rm s}\,\delta\rho/\rho $ (\citealt{gaspari13}, Eq.~22). Note that our average temperatures are spectroscopic emission-weighted temperatures \citep[e.g.][]{mazzotta04}, and thus our estimation of the sound speed does not exactly match the mean mass-weighted sound speed in the system. In Figure \ref{f:a2d_data}, we show the main result of the current study plotting the cluster 1.4~GHz radio power versus the estimated turbulent velocity dispersion. Again, a segregation is observed between the radio halo and upper limit populations. For radio halo clusters, the radio power appears to correlate with the velocity dispersion, with Pearson coefficient $\rho=0.80\pm0.02$ for 25 data points, indicating that the two quantities are significantly correlated. 

To describe the relation between $\sigma_v$ and $P_{1.4}$, we fitted the data with a power law. Namely, we modeled the data as 

\begin{equation}\log \left(\frac{P_{\rm 1.4 GHz}}{\rm 10^{24}\, W\, Hz^{-1}}\right) = \log P_{0} + \alpha \log\left(\frac{\sigma_v}{\rm 500\, km\, s^{-1}} \right).\end{equation}

We used the Bayesian routine \texttt{linmix\_err} \citep{kelly07} to fit the data, taking both the data points and radio upper limits into account. The best-fitting values for the parameters are $\alpha=3.27_{-0.61}^{+0.71}$, $P_{0}=2.34_{-0.49}^{+0.53}$, with an intrinsic scatter $\sigma_{\ln P|\sigma_v}=0.44_{-0.13}^{+0.18}$. We note that the relation between the projected surface-brightness fluctuations and three-dimensional velocity field should exhibit a substantial scatter, thus the scatter of $\sim40\%$ observed here is not surprising.

\section{Discussion} \label{s:interp}

\noindent

The main results of this work have potentially important implications on the origin of radio halos, which we discuss here. While previous works have found that radio halos arise almost only in morphologically disturbed systems \citep{buote01,cassano10,cassano13,cuciti15}, as evidenced e.g. by their centroid shifts, our analysis shows that radio halo clusters remain on average more perturbed even after subtracting the large-scale gas distribution and masking substructures. We expect that surface-brightness fluctuations on top of the large-scale gas distribution trace the presence of residual gas motions. Numerical simulations predict that ICM turbulence should be the dominant source of residual motions and hence of ICM fluctuations \citep[e.g.][]{vazza09b,Lau:2009}, although other sources of fluctuations (e.g. shocks, cold fronts, ram-pressure stripping) play a role to some extent. Our work thus establishes a connection between turbulent motions and particle acceleration in galaxy clusters, which corroborates a key prediction of the turbulent re-acceleration scenario \citep[e.g.][]{brunetti01}.

In case the density fluctuations measured here can be entirely ascribed to turbulence, Fig. \ref{f:a2d_data} shows that the recovered turbulent velocity dispersion correlates with the radio power at 1.4 GHz, which can be interpreted in the framework of the turbulent re-acceleration scheme. The energy rate per unit volume related to turbulence can be estimated as
\begin{equation} \label{e:Pturb}
P_{\rm turb} \approx  9.8\times10^{-25} \left(\frac{\sigma_{v}}{500\,\kms}\right)^3\left(\frac{n_{\rm gas}}{10^{-2}\rm \,cm^{-3}}\right) \left(\frac{L_{\rm inj}}{\rm 500\,kpc}\right)^{-1} {\rm erg\,s^{-1}\,cm^{-3}}
\end{equation}
In the classical Kolmogorov model, the kinetic power across the cascade is preserved, regardless of the specific dissipation mechanism. While the specific scales and mechanisms for the ultimate dissipation of turbulent motions in the ICM are currently unknown \citep[see e.g.][]{brunetti11}, it is reasonable to assume that a fraction of the kinetic power through the cascade goes into the re-acceleration of radio emitting electrons. The exact relation between $P_{\rm radio}$ and $P_{\rm turb}$ is uncertain as it depends on many unknown factors. Yet our data support (within $1\sigma$) a simple $P_{\rm radio}\propto \sigma_{v}^3\propto P_{\rm turb}$, suggesting that $A_{2D}$ can capture the bulk of the energy involved in the re-acceleration process. 

Although the segregation between radio halo and radio quiet clusters in Fig. \ref{fig:distribution} and \ref{f:a2d_data} is clear, there are obviously a number of outliers. Among these, we note the cases of A209 and A781. A209 is the radio halo cluster with the lowest value of $M_{\rm 1D}$ ($0.13\pm0.01$) and $\sigma_v$ ($273\pm15$ km/s). Such values would classify this object firmly within the ``upper limit'' quadrant. A209 could be a fading radio halo in which turbulence is now being dissipated. Alternatively, projection effects may hide the presence of large-scale fluctuations, which would lead to an underestimate of the turbulent Mach number. Conversely, A781 exhibits a fairly high Mach number $M_{\rm 1D}=0.22\pm0.02$ and $\sigma_v=457\pm42$, yet this system does not host a radio halo \citep{venturi11}. One possibility is that this system is in the early stage of a merger and that the turbulent cascade is not fully developed, such that the onset of particle acceleration at the micro scale has not yet taken place. 

As a word of caution, we note that the conversion between 2D amplitude and turbulent velocity dispersion assumes that all the observed fluctuations are induced by turbulent motions. The presence of additional perturbing phenomena such as unresolved gas clumps or shock fronts will lead to an increase in the fractional amplitude of density perturbations. Although deviating regions such as clear substructures were masked during this analysis, the values reported here might overestimate the intrinsic turbulent Mach number.

Future X-ray imaging spectrometers such as \emph{ATHENA} \citep{Nandra:2013} or the {\it Hitomi} recovery mission \emph{XARM} will allow us to measure the velocity dispersion directly and will set strong constraints on the models describing the acceleration of relativistic electrons by turbulence in the ICM.

\section{Conclusions} \label{s:conc}
\noindent
In this Letter, we presented measurements of the projected amplitude of gas density fluctuations in the ICM at a single scale $k^{-1}=660$ kpc and, for the first time, correlated it with the presence or not of radio emission for a sample of 51 galaxy clusters. Our results can be summarized as follows: 

\begin{itemize}
\item We observe a clear segregation between the level of ICM fluctuations between the clusters exhibiting a radio halo and the ones where no radio emission has been detected. The difference between the two populations is significant at the $4.6\sigma$ level. 

\item If the measured fluctuations are interpreted as being entirely due to the presence of residual gas motions, the 1D Mach number of turbulent motions is found to be larger in the radio halo populations by about a factor of two compared to the clusters where no radio emission was detected.

\item The turbulent velocity dispersion $\sigma_v=M_{\rm 3D}\,c_{\rm s}$ of radio halo systems correlates with the observed radio power (correlation coefficient $0.80\pm0.02$). The best-fit relation reads $P_{1.4}\propto \sigma_v^{3.3\pm0.7}$ with 44\% scatter. Thus, the radio power is nearly proportional to the turbulent energy rate $P_{\rm turb}\propto\sigma_v^3$.

\item Provided that surface-brightness fluctuations are probing gas motions in the ICM, our results corroborate stochastic acceleration via turbulence as the most likely mechanism to boost the emergence of radio halos in galaxy clusters. 
\end{itemize}

\section*{Acknowledgements}
\noindent
This work was supported in part by NASA Chandra grant GO7-18121X.
M.G. is supported by NASA through Einstein Postdoctoral Fellowship Award Number PF5-160137 issued by the Chandra X-ray Observatory Center, which is operated by the SAO for and on behalf of NASA under contract NAS8-03060. F.V. acknowledges financial support from the EU Horizon 2020 research and innovation program under the Marie-Sklodowska-Curie grant agreement no.~664931. S.E. acknowledges the financial support from contracts ASI-INAF I/009/10/0, NARO15 ASI-INAF I/037/12/0 and ASI 2015-046-R.0.

\begin{center}
\begin{deluxetable*}{lccccccc}
\tabletypesize{\small}
\tablecaption{\label{tab:prop}Properties of the cluster sample\tablenotemark{1}
}
{\setstretch{0.7}
\tablehead{\colhead{Cluster} & \colhead{z} &\colhead{$kT_{\rm 1 Mpc}$ [keV]} &\colhead{$P_{1.4}$ [W/Hz]} &  \colhead{$A_{2D}$} &  \colhead{Mach$_{\rm 1D}$} & \colhead{$\sigma_{v}$ [km s$^{-1}$]} & \colhead{Instrument}}
\tablecolumns{5}
\startdata
Upper Limits \\ 
A2697 & 0.232 & $ 6.94 \pm 0.14 $ & $< 0.41 $ & $ 0.06 \pm 0.01 $ & $ 0.12 \pm 0.01 $ & $ 250 \pm 27 $ & X \\ 
A141 & 0.23 & $ 5.9 \pm 0.15 $ & $< 0.43 $ & $ 0.08 \pm 0.01 $ & $ 0.16 \pm 0.02 $ & $ 325 \pm 38 $ & X \\ 
A3088 & 0.2537 & $ 6.42 \pm 0.16 $ & $< 0.43 $ & $ 0.07 \pm 0.01 $ & $ 0.14 \pm 0.02 $ & $ 291 \pm 39 $ & X \\ 
RXCJ0437.1+0043 & 0.285 & $ 6.45 \pm 0.2 $ & $< 0.65 $ & $ 0.04 \pm 0.01 $ & $ 0.09 \pm 0.02 $ & $ 179 \pm 39 $ & X \\ 
RXCJ1115.8+0129 & 0.3499 & $ 6.21 \pm 0.11 $ & $< 0.47 $ & $ 0.06 \pm 0.01 $ & $ 0.11 \pm 0.02 $ & $ 227 \pm 41 $ & X \\ 
A2631 & 0.2779 & $ 7.61 \pm 0.3 $ & $< 0.41 $ & $ 0.09 \pm 0.01 $ & $ 0.16 \pm 0.02 $ & $ 374 \pm 48 $ & X \\ 
A2645 & 0.251 & $ 6.28 \pm 0.18 $ & $< 0.59 $ & $ 0.08 \pm 0.01 $ & $ 0.15 \pm 0.01 $ & $ 315 \pm 25 $ & X \\ 
A2667 & 0.2264 & $ 5.99 \pm 0.08 $ & $< 0.45 $ & $ 0.08 \pm 0.01 $ & $ 0.15 \pm 0.01 $ & $ 310 \pm 24 $ & X \\ 
Z348 & 0.2537 & $ 2.94 \pm 0.05 $ & $< 0.65 $ & $ 0.05 \pm 0.01 $ & $ 0.10 \pm 0.02 $ & $ 144 \pm 22 $ & X \\ 
RXJ0142.0+2131 & 0.2803 & $ 7.37 \pm 0.58 $ & $< 0.45 $ & $ 0.09 \pm 0.02 $ & $ 0.18 \pm 0.03 $ & $ 402 \pm 70 $ & C \\ 
A267 & 0.231 & $ 5.64 \pm 0.17 $ & $< 0.34 $ & $ 0.08 \pm 0.01 $ & $ 0.16 \pm 0.01 $ & $ 306 \pm 27 $ & X \\ 
RXJ0439.0+0715 & 0.23 & $ 6.23 \pm 0.48 $ & $< 0.46 $ & $ 0.07 \pm 0.02 $ & $ 0.13 \pm 0.04 $ & $ 272 \pm 77 $ & C \\ 
RXJ0439.0+0520 & 0.208 & $ 6.53 \pm 0.25 $ & $< 0.32 $ & $ 0.01 \pm 0.04 $ & $ 0.02 \pm 0.08 $ & $ 41 \pm 167 $ & C \\ 
A611 & 0.288 & $ 6.63 \pm 0.45 $ & $< 0.43 $ & $ 0.07 \pm 0.03 $ & $ 0.13 \pm 0.05 $ & $ 273 \pm 105 $ & X \\ 
Z2089 & 0.2347 & $ 3.82 \pm 0.13 $ & $< 0.26 $ & $ 0.07 \pm 0.01 $ & $ 0.13 \pm 0.01 $ & $ 205 \pm 16 $ & X \\ 
A781 & 0.2984 & $ 6.21 \pm 0.12 $ & $< 0.36 $ & $ 0.11 \pm 0.01 $ & $ 0.22 \pm 0.02 $ & $ 457 \pm 42 $ & X \\ 
Z2701 & 0.214 & $ 5.41 \pm 0.22 $ & $< 0.35 $ & $ 0.07 \pm 0.01 $ & $ 0.13 \pm 0.02 $ & $ 250 \pm 42 $ & C \\ 
A1423 & 0.213 & $ 7.65 \pm 0.39 $ & $< 0.38 $ & $ 0.07 \pm 0.01 $ & $ 0.13 \pm 0.01 $ & $ 306 \pm 32 $ & C \\ 
A1576 & 0.279 & $ 7.73 \pm 0.46 $ & $< 0.64 $ & $ 0.10 \pm 0.02 $ & $ 0.19 \pm 0.03 $ & $ 429 \pm 69 $ & C \\ 
RXJ1532.9+3021 & 0.345 & $ 4.89 \pm 0.07 $ & $< 0.66 $ & $ 0.04 \pm 0.00 $ & $ 0.07 \pm 0.01 $ & $ 128 \pm 15 $ & X \\ 
A2146 & 0.2343 & $ 6.59 \pm 0.17 $ & $< 0.39 $ & $ 0.08 \pm 0.01 $ & $ 0.15 \pm 0.02 $ & $ 319 \pm 35 $ & C \\ 
A2261 & 0.224 & $ 8.01 \pm 0.31 $ & $< 0.32 $ & $ 0.08 \pm 0.01 $ & $ 0.15 \pm 0.02 $ & $ 357 \pm 42 $ & X \\ 
A2537 & 0.2966 & $ 7.46 \pm 0.34 $ & $< 0.51 $ & $ 0.06 \pm 0.02 $ & $ 0.12 \pm 0.03 $ & $ 263 \pm 78 $ & X \\ 
RXJ0027.6+2616 & 0.3649 & $ 5.29 \pm 0.58 $ & $< 0.74 $ & $ 0.08 \pm 0.01 $ & $ 0.15 \pm 0.03 $ & $ 284 \pm 53 $ & X \\ 
Z5768 & 0.266 & $ 3.34 \pm 0.23 $ & $< 0.36 $ & $ 0.09 \pm 0.01 $ & $ 0.17 \pm 0.02 $ & $ 260 \pm 35 $ & X \\ 
S780 & 0.2357 & $ 6.67 \pm 0.22 $ & $< 0.38 $ & $ 0.09 \pm 0.01 $ & $ 0.17 \pm 0.01 $ & $ 354 \pm 28 $ & C \\ 
Radio Halos \\ 
A2744 & 0.307 & $ 8.69 \pm 0.32 $ & $ 18.62 \pm 0.94 $ & $ 0.14 \pm 0.02 $ & $ 0.26 \pm 0.04 $ & $ 629 \pm 89 $ & C \\ 
A209 & 0.206 & $ 6.67 \pm 0.12 $ & $ 1.99 \pm 0.21 $ & $ 0.07 \pm 0.00 $ & $ 0.13 \pm 0.01 $ & $ 273 \pm 15 $ & X \\ 
A2163 & 0.203 & $ 15.22 \pm 0.16 $ & $ 22.91 \pm 1.16 $ & $ 0.12 \pm 0.01 $ & $ 0.22 \pm 0.02 $ & $ 711 \pm 65 $ & X \\ 
RXCJ2003.5-2323 & 0.3171 & $ 9.35 \pm 0.53 $ & $ 10.71 \pm 1.73 $ & $ 0.13 \pm 0.01 $ & $ 0.24 \pm 0.02 $ & $ 603 \pm 45 $ & C \\ 
A520 & 0.199 & $ 7.21 \pm 0.16 $ & $ 2.45 \pm 0.18 $ & $ 0.13 \pm 0.01 $ & $ 0.24 \pm 0.02 $ & $ 536 \pm 55 $ & C \\ 
A773 & 0.217 & $ 7.65 \pm 0.19 $ & $ 1.48 \pm 0.16 $ & $ 0.10 \pm 0.01 $ & $ 0.20 \pm 0.02 $ & $ 451 \pm 35 $ & X \\ 
A1758a & 0.28 & $ 7.22 \pm 0.17 $ & $ 5.75 \pm 0.98 $ & $ 0.14 \pm 0.02 $ & $ 0.27 \pm 0.03 $ & $ 589 \pm 65 $ & C \\ 
A2219 & 0.2281 & $ 10.02 \pm 0.25 $ & $ 5.63 \pm 0.8 $ & $ 0.09 \pm 0.01 $ & $ 0.17 \pm 0.02 $ & $ 432 \pm 40 $ & X \\ 
A521 & 0.2475 & $ 6.41 \pm 0.21 $ & $ 1.45 \pm 0.13 $ & $ 0.12 \pm 0.01 $ & $ 0.22 \pm 0.02 $ & $ 462 \pm 44 $ & C \\ 
A697 & 0.282 & $ 9.66 \pm 0.85 $ & $ 1.51 \pm 0.14 $ & $ 0.09 \pm 0.02 $ & $ 0.18 \pm 0.03 $ & $ 458 \pm 81 $ & X \\ 
A1300 & 0.3075 & $ 7.85 \pm 0.25 $ & $ 3.8 \pm 1.43 $ & $ 0.14 \pm 0.01 $ & $ 0.26 \pm 0.03 $ & $ 595 \pm 62 $ & C \\ 
CL0016+16 & 0.541 & $ 9.42 \pm 0.3 $ & $ 5.01 \pm 0.31 $ & $ 0.10 \pm 0.01 $ & $ 0.19 \pm 0.01 $ & $ 472 \pm 35 $ & C \\ 
A1914 & 0.1712 & $ 11.38 \pm 0.67 $ & $ 5.62 \pm 0.43 $ & $ 0.12 \pm 0.01 $ & $ 0.24 \pm 0.02 $ & $ 657 \pm 53 $ & C \\ 
A665 & 0.1819 & $ 7.53 \pm 0.17 $ & $ 2.51 \pm 0.21 $ & $ 0.12 \pm 0.01 $ & $ 0.24 \pm 0.02 $ & $ 539 \pm 39 $ & X \\ 
A545 & 0.154 & $ 6.57 \pm 0.09 $ & $ 1.41 \pm 0.22 $ & $ 0.10 \pm 0.01 $ & $ 0.19 \pm 0.01 $ & $ 399 \pm 24 $ & X \\ 
Coma & 0.0231 & $ 8.28 \pm 0.13 $ & $ 0.76 \pm 0.06 $ & $ 0.09 \pm 0.01 $ & $ 0.18 \pm 0.02 $ & $ 420 \pm 45 $ & R \\ 
A2256 & 0.0581 & $ 7.65 \pm 0.63 $ & $ 0.85 \pm 0.08 $ & $ 0.13 \pm 0.01 $ & $ 0.25 \pm 0.02 $ & $ 571 \pm 56 $ & R \\ 
Bullet & 0.296 & $ 14.58 \pm 0.4 $ & $ 23.44 \pm 1.51 $ & $ 0.17 \pm 0.02 $ & $ 0.31 \pm 0.03 $ & $ 977 \pm 99 $ & C \\ 
A2255 & 0.0806 & $ 5.81 \pm 0.2 $ & $ 0.81 \pm 0.17 $ & $ 0.09 \pm 0.01 $ & $ 0.18 \pm 0.01 $ & $ 365 \pm 22 $ & R \\ 
A2319 & 0.0557 & $ 9.6 \pm 0.3 $ & $ 2.45 \pm 0.19 $ & $ 0.11 \pm 0.01 $ & $ 0.21 \pm 0.02 $ & $ 529 \pm 46 $ & R \\ 
MACSJ0717.5+3745 & 0.548 & $ 13.59 \pm 0.68 $ & $ 52.48 \pm 20.56 $ & $ 0.21 \pm 0.02 $ & $ 0.40 \pm 0.03 $ & $ 1206 \pm 96 $ & C \\ 
A1995 & 0.3186 & $ 7.58 \pm 0.41 $ & $ 1.66 \pm 0.23 $ & $ 0.08 \pm 0.01 $ & $ 0.14 \pm 0.03 $ & $ 326 \pm 59 $ & C \\ 
MACSJ1149.5+2223 & 0.544 & $ 13.14 \pm 0.9 $ & $ 2.29 \pm 0.95 $ & $ 0.15 \pm 0.02 $ & $ 0.29 \pm 0.03 $ & $ 853 \pm 94 $ & C \\ 
PLCKG171.9-40.7 & 0.27 & $ 12.78 \pm 0.79 $ & $ 4.9 \pm 1.35 $ & $ 0.08 \pm 0.01 $ & $ 0.15 \pm 0.01 $ & $ 438 \pm 40 $ & X \\ 
A754 & 0.0542 & $ 8.33 \pm 0.05 $ & $ 0.63 \pm 0.07 $ & $ 0.15 \pm 0.01 $ & $ 0.28 \pm 0.02 $ & $ 676 \pm 46 $ & R \\ 
\enddata
}
\tablenotetext{1}{\emph{a)} Cluster name; \emph{b)} Redshift; \emph{c)} Spectroscopic temperature extracted within a circular region of 1 Mpc radius; \emph{d)} Radio power at 1.4 GHz in units of $10^{24}$ W/Hz \citep[from][]{cassano13}; \emph{e)} Amplitude of two-dimensional X-ray brightness fluctuations; \emph{f)} 1D Mach number $M_{\rm 1D}=2.3\,\delta\rho/\rho$; \emph{g)} Turbulent velocity dispersion $\sigma_v=M_{\rm 3D}\,c_{\rm s}$, with $\,c_{\rm s}\approx(\gamma kT_{\rm 1 Mpc}/\mu m_p)^{0.5}$; \emph{h)} X-ray instrument used for the analysis (X=\emph{XMM-Newton}, C=\emph{Chandra}, R=\emph{ROSAT}).}
\end{deluxetable*}
\end{center}


\end{document}